# A Thermal Acid Calcification Cause for Seasonal Oscillations in the Increasing Keeling Curve

Ivan R. Kennedy[1]*, John Runcie[2], Angus N. Crossan[3], Raymond Ritchie[4], Jennifer Marohasy[5]

[1] School of Life and Environmental Sciences, University of Sydney Association of Professors NSW 2006, Australia.
[2] Aquation Pty Ltd., P.O. Box 3146, Woy Woy, NSW 2257, Australia.
[3] Quick Test Technologies Pty Ltd., www.ackle.au, Australia.
[4] Faculty of Technology and Environment, Prince Songkla University, Phuket 83120, Thailand.
[5] ClimateLab Pty Ltd, PO Box 170, Yeppoon, QLD 4703, Australia.

*Corresponding author: ivan.kennedy@sydney.edu.au

**Abstract:** Why do atmospheric carbon dioxide levels rise and fall seasonally measured on Mauna Loa? This study explores the thermal acid-calcification (TAC) hypothesis, suggesting that seasonal temperature shifts in surface seawater trigger acid pH-driven $CO_2$ emissions caused by calcification. Using oceanographic data, we modeled how temperature affects dissolved inorganic carbon including $CO_2$, bicarbonate, and carbonate. Our findings reveal that warming waters absorb atmospheric CO2 by promoting calcium carbonate formation, acidifying seawater and boosting $CO_2$ release to the atmosphere in late autumn and winter, when atmospheric $CO_2$ becomes highest. The model predicts a net annual $CO_2$ rise of 2 ppmv, driven by calcification rather than land-based processes. Seasonal pH swings of 0.04 units corroborate this mechanism. The TAC hypothesis indicates that continued ocean warming, not just fossil fuels, contribute to rising $CO_2$ levels, calling for deeper investigation into marine carbon dynamics.

## Introduction

Carbon dioxide is everywhere. Its fugacity ensures that it will penetrate every space to minimise its chemical potential. Children playing use it with baking soda to simulate volcanoes, adding excess vinegar to acidify the sodium carbonate solution thus releasing $CO_2$ at high pressure, expanding the foam. Beverages including mineral water or beer are supercharged with it, to provide sparkling reactions on the palate, reducing the pressure under seal by escaping freely into air at a mean relative volume close to 420 ppm.

By contrast, in any closed space containing liquid water the gas space will eventually reach an equilibrium $pCO_2$ value where the chemical potential of $CO_2$ in the liquid and gas phases become equal, a function of the pH value of the aqueous phase and concentrations of dissolved inorganic carbon (DIC). Add vinegar to the liquid and the $pCO_2$ will rise accordingly, reducing the concentrations of carbonate and bicarbonate in the solution as pH falls. For inorganic and environmental chemists, a question should arise regarding the equilibrium for $CO_2$ distribution between global surface water on lands and seas with the atmosphere. Does such a pH equilibrium, as the outcome of a statistical process in many localities, exist globally? We predict that evidence favoring this hypothesis will soon be accepted as an emerging environmental principle.

**The Keeling Curve for atmospheric $p$CO$_2$ in parts per million by volume**

Fossil fuel emissions of CO$_2$ mix into air, some distributing into surface waters reaching Henry Law or coefficient distributions as a function of $p$CO$_2$, dissolved inorganic concentrations of bicarbonate, carbonate, and pH. Could the rising Keeling $p$CO$_2$ curve as first measured by Charles Keeling from 1958 be such a quasi-equilibrium on the Big Island of Hawaii, Mauna Loa? Unable to make reproducible measurements of CO$_2$ on land, Keeling sought sampling locations with more stability. High up the volcanic cone of Mauna Loa at 3700 m protected from volcanic emissions was one such location.

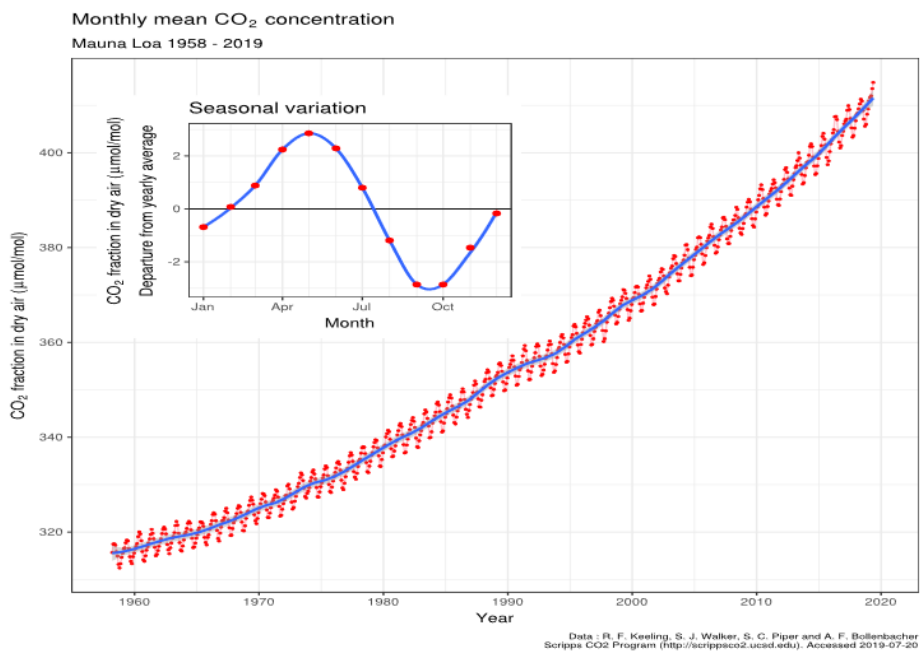

**Fig. 1.** The Keeling curve of atmospheric CO$_2$ partial pressure at 3200 m on Mauna Loa, Hawaii. Data from Dr. Pieter Tans, NOAA/ESRL and Dr.Ralph Keeling, Scripps Institution of Oceanography.CC BY-SA 4.0,https://commons.wikimedia.org/w/index.php?curid=40636957

The very stuff of plant life on Earth in photosynthesis as well as in the structural basis of all living creatures, we are told by the United Nations Intergovernmental Panel on Climate Change (IPCC) (*2*) that the continuing rise of CO$_2$ in the Keeling curve shown in Figure 1 now threatens global catastrophe from global warming. Such a paradoxical contrast for good and bad challenges credibility, given the longevity of life on Earth.

**Transfers of CO$_2$ out of the ocean surface in winter versus that entering in summer**

We propose that a quasi-equilibrium exists between a falling pH value in surface water, favoring CO$_2$.emisssion. Falling pH values in the surface water of the oceans have been an enigma, invisible to scientific view until recently after the year 2000. Our logic is supported in our articles (*3, 4*) where we describe the basis for the thermal acid-calcification (TAC) hypothesis, also using data cited from others. Acidic calcification is thermodynamically favored in warming surface seawater, particularly in northern oceans in spring and summer with shallow mixing zones and higher temperature ranges. This raises the fugacity or potential pressure of CO$_2$ in seawater to its peak value in summer when the $pCO_2$ in air is minimal, causing its forceful emission into air in the next autumn reaching a maximum $pCO_2$ in late winter (Fig. 1, seasonal variation insert). This process of calcification and accelerated $CO_2$ emission and its incomplete absorption in spring is a probable cause of the seasonal oscillation of $pCO_2$ in the Keeling curve on Mauna Loa, Hawaii, rather than imbalances between rates of terrestrial photosynthesis and respiration, an unproven assumption. So, is warming of surface seawater a result or a cause of the increasing $pCO_2$ in the atmosphere?

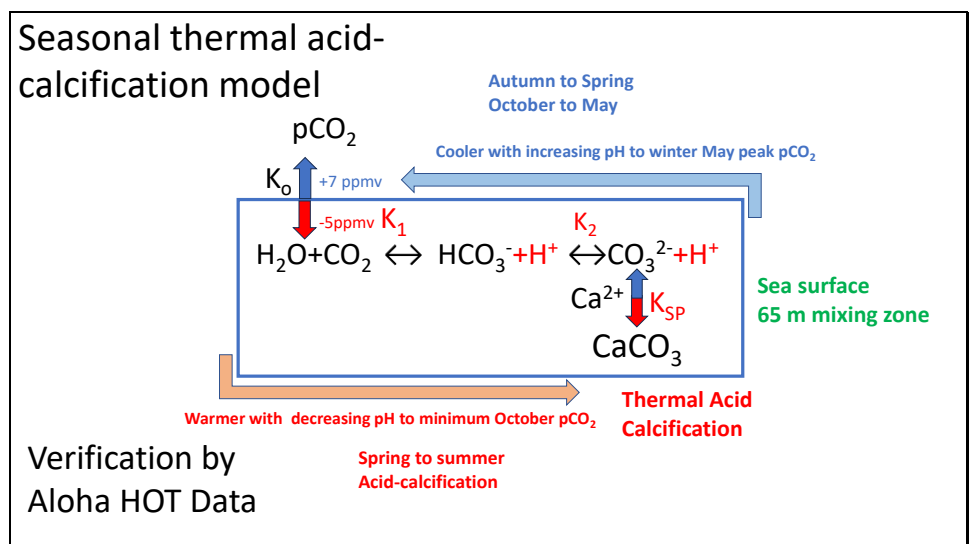

**Fig. 2.** Thermal acid-calcification model for seasonal and longer-term generation of the Keeling curve. The Thermal model (*3*), showed calcification is favored by increase in temperature giving decreasing pH values in summer, reversing in cooler conditions to more alkaline pH in winter. Note that precipitation of $CaCO_3$ in spring to summer removes $CO_3^{2-}$ that is replaced from $HCO_3^-$ with more acidity, provided by absorption of $CO_2$ from air up to October. However, as the pH falls the steady state concentration of [CO2} increases favoring photosynthesis.

**A testable thermal acid-calcification hypothesis**

Figure 2 shows our hypothesis for $CO_2$ cycling in the global ecosystem. To many it is counterintuitive that $CaCO_3$ is less soluble in warmer water. However, this is a scientific fact consistent with $CO_2$ being absorbed from the atmosphere into surface seawater, as shown in Figure 2 in spring and summer, to form calcite. The solubility of calcium sulphate also decreases as temperature increases. This behavior known as retrograde solubility is uncommon: dissolution of most of salts increases with temperature. The retrograde solubility of calcium sulphate is also responsible for its precipitation in the hottest zones of heating systems and for its contribution to the formation of scale in boilers together with the precipitation of calcium carbonate whose solubility also decreases when $CO_2$ outgasses from hot water or can escape out of the system (https://en.wikipedia.org/wiki/Calcium_sulfate). Our conclusion from this anomalous solubility behavior is that all the current fossil emissions could now be consumed as a net process after mixing in air, either by greening photosynthesis or even in the ocean in mixing layers at the surface. Even cold areas conduct the processes shown by forms of $CaCO_3$ with lower solubility. The scale of the seawater processes modelled is sufficient to explain the seasonal changes as dominated by variations in temperature.

This thermal model offers a plausible alternative mechanism for the increasing pressure $p$CO$_2$ in ppm by volume each year, a process of dynamic equilibration of the declining pH value of surface seawater caused by calcification, this is necessary to support high rates of photosynthesis by phytoplankton. This hypothesis is consistent with the absorption of about half the fossil emissions of CO$_2$ in surface seawater with the remainder contributing to the greening of the Earth caused by enhanced photosynthesis, shown by spectral observations from satellites. Current IPCC policies for zero carbon emissions such as renewable sources of energy must fail, if this hypothesis is confirmed by year-by-year increasing depositions of CaCO$_3$ from surface seawaters; this would be about two-thirds of a mole per square meter of the Earth's surface, or 10 µmoles of calcite per kg of surface seawater annually, producing an increase of 2 ppmv in $p$CO$_2$.

A feature of the annual increase in $p$CO$_2$ as ppmv by volume, (not weight) is the seasonal oscillation (Fig, 1). The pressure of CO$_2$ in the northern hemisphere peaks in late April and reaches its minimum in October. Keeling's interpretation was that terrestrial photosynthesis extracted CO$_2$ from the atmosphere in spring and early summer and that respiration of organic carbon forming CO$_2$ predominated in autumn and winter. This is true of the day and night cycle and daily over a wheat crop when growing vegetatively. However, is it true regionally over the vast Pacific Ocean near Hawaii? We suggest there is no such evidence.

We investigated this question using modeling (*3, 4*) and extensive data gathered by oceanographers (*5, 6*), together with their excellent algorithms for finding the effects of seasonal variation of temperature and salt concentration on equilibrium constants. These include the Henry coefficient K$_o$, the reaction of CO$_2$ and water to form bicarbonate (K$_1$), and then carbonate (K$_2$), as well as the solubility product for different forms of calcium carbonate (CaCO$_3$), K$_{sp}$. The Henry coefficient is independent of pH, although K$_1$ and K$_2$ clearly are, regulating the distribution of DIC between carbonate, bicarbonate and CO$_2$ in solution All these reactions are reversible and two involve either the release or the absorption of acidity as hydrogen ions (H$^+$) (Fig. 2). Although we usually teach that carbonic acid (H$_2$CO$_3$) exists, this is so unstable in water that it hardly needs consideration.

**Acid calcification is essential for phytoplankton**

Any process of strong acidification of surface seawater will raise the concentration of carbon dioxide as [CO$_2$] available to phytoplankton for photosynthesis. Bicarbonate cannot be a direct substrate for photosynthesis although the presence of the enzyme carbonic anhydrase speeds up its interconversion with CO$_2$. Our published modeling analysis confirmed that CaCO$_3$ precipitation is strongly favored by warmer temperatures (Table 1). Indeed, all the reaction equilibria in seawater are displaced to the right in Figure 2 acidifying the water, although the equilibrium between CO$_2$ concentration and $p$CO$_2$ in air favors a lower concentration [CO$_2$} in water in summer, compared to winter, when it is greatest. Our results even confirmed that the formation of CaCO$_3$ as calcite is predicted to increase in summer as water becomes warmer (Table 1b).

 Thus, we can expect more limestone formation in summer if the carbonate concentration reaches a sufficient level, favored by added warming. The decline in average pH values in surface seawater to about 8.05 from pH 8.20 could explain the increased $p$CO$_2$ in the atmosphere of 140 ppmv since 1800 as a matter of dynamic equilibrium. Caused by calcification, this would require a simultaneous equivalent deposition of limestone as sediment, though only an increase of about 10 µmoles per kg of surface seawater, or a net 1 mg per kg each year. This is a key prediction for experimental testing of the TAC hypothesis.

An important feature of $CaCO_3$ formation is acidification; removal of $CO_3^{2-}$ ions by precipitation displaces the equilibrium to generate even more acidification than that from dissolving $CO_2$ alone in forming bicarbonate and carbonate. Indeed, each precipitation of a µmole of carbonate will reduce pH values by an equivalent amount. Even if calcite is not visibly formed, the fact that much of the surface layer in oceans are measured to be supersaturated with $CO_3^{2-}$ ions suggests that the activity of $Ca^{2+}$ ions is diminished by binding to $CO_3^{2-}$ ions when warmer, already regarded as low as only 0.20 in seawater. Warmer water also dissociates water clusters, diminishing the shielding of $Ca^{2+}$ ions from negatively charged carbonate and thus promoting acidification from bicarbonate dissociation to carbonate.

This process of calcification is essential for photosynthetic life in the ocean since it actually raises the concentration of $CO_2$, equivalent to a fugacity equal to 7-10 ppmv $pCO_2$ in air for each fall of 0.01 pH units, an output of our thermal model; we now call this the acid-calcification thermal (TAC) hypothesis. The $K_m$ concentration value for the enzyme absorbing $CO_2$ known as Rubisco, the most common protein on Earth, is about the same as that of the concentration $[CO_2]$ in air and seawater, so any increased value speeds up photosynthesis. Without calcification, nearly all photosynthetic life in seawater would cease for lack of sufficient $CO_2$ for Rubisco from bicarbonate (7).

**Le Chatelier's principle favors calcification when seawater is warmer and decalcification when cooler**

Changes with temperature (3) in the equilibrium constants for DIC of Figure 2 are shown in Table 1A, estimated using the excellent algorithms of Millero and many colleagues. These algorithms, summarized in the Emerson and Hedges treatise, *Chemical Oceanography and Marine Carbon Cycling* (5), are excellent for analysis of seawater samples, but rarely used in modeling. The Henry coefficient ($K_o$) is almost halved by over a 20 °C increase in temperature. The pK values near pH 6 and 9 where $[CO_2]$ and $[HCO_3^-]$ or $[HCO_3^-]$ and $[CO_3^{2-}]$ respectively are of equal concentration both decline with temperature at a similar rate, showing $K_1$ and $K_2$ values increase (i.e., more acid). The solubility product shows a maximum near 285 K, being less soluble both warmer and colder than this temperature.

**Table 1.** (A) Outputs of estimated variation with temperature of equilibrium constants used for modeling of surface seawater at $pCO_2$air 420 ppmv

| Temperature K | C | $K_o=Z_{sw}$ | $pK_1$ | $pK_2$ | $K_w$ x$10^{14}$ | $pK_w$ | $K_{sp}$ x$10^{-7}$ $[Ca^{2+}][CO_3^{2-}]$ | µmolar $CO_2$ in seawater |
|---|---|---|---|---|---|---|---|---|
| 278.15 | 5 | 0.05213 | 6.042 | 9.292 | 0.855 | 14.07 | 4.309 | 21.90 |
| 283.15 | 10 | 0.04388 | 5.984 | 9.203 | 1.443 | 13.84 | 4.317 | 18.43 |
| 288.15 | 15 | 0.03746 | 5.931 | 9.118 | 2.380 | 13.62 | 4.315 | 15.73 |
| 293.15 | 20 | 0.03241 | 5.882 | 9.035 | 3.839 | 13.42 | 4.300 | 13.61 |
| 298.15 | 25 | 0.02839 | 5.837 | 8.955 | 6.063 | 13.22 | 4.272 | 11.92 |

Results varying with temperature estimated with Emerson polynomials (3, 5)

1(B) Van't Hoff equation outputs of standard enthalpy, Gibbs energy and entropy changes for the $K_1$ and $K_2$ equilibria

| Temperature K | C | $pK_1$ | ΔH° kJ/mol | ΔG° kJ/mol | ΔS° kJ/K | $pK_2$ | ΔH° kJ/mol | ΔG° kJ/mol | ΔS° kJ/K/mol |
|---|---|---|---|---|---|---|---|---|---|
| 278.15 | 5 | 6.042 | 17.490 | 32.301 | -0.053 | 9.292 | 26.837 | 49.672 | -0.081 |
| 283.15 | 10 | 5.984 | | | | 9.203 | | | |
| 288.15 | 15 | 5.931 | 16.273 | 32.698 | -0.057 | 9.118 | 26.751 | 50.152 | -0.081 |
| 293.15 | 20 | 5.882 | | | | 9.035 | | | |
| 298.15 | 25 | 5.837 | 15.059 | 33.160 | -0.061 | 8.955 | 26.771 | 50.900 | -0.082 |

Estimates made with pK results from (a) using programmable machine language software (*3*)

Similar endothermic results were obtained for the calcification reaction for precipitation of calcite, which should also apply to supersaturation conditions.

**Van't Hoff Equation for Changes in Temperature**

$$ln\left(\frac{K2}{K1}\right) = \Delta H/R(1/T_1 - 1/T_2)$$

Van't Hoff analysis (Table 1B) to measure changes in equilibrium constants ($K_2/K_1$) from variations in enthalpy (Δ*H*) showed that all reactions favored in summer are endothermic with ΔH° positive, absorbing heat at higher temperatures, cooling surface seawater. In winter when the reverse reactions are favored there will be a tendency to warm seawater slightly, these thermal effects being good demonstrations of Le Chatelier's principle. When environmental conditions force reaction in one direction, natural processes oppose such changes.

**Data from Others Helps Solve the Seasonal Enigma**

Other evidence from seawater samplings by Dore et al. (*8*) showing a year-by-year decrease in pH values of about 0.003 units, with a much larger seasonal variation in seawater. The ALOHA or HOT data in particular shows this seasonal variation in pH values of 0.04 units in seawater, with matching between minimum pH and maximum fugacity of $CO_2$ in surface seawater (Fig. 3). Our modeling indicates that absorption of $CO_2$ alone is unable to explain the decreased pH value abiotically from this cause. Significant calcification removing more $CO_2$ from the atmosphere is needed to explain the additional [$H^+$] increase and pH changes involved. We regard this as consistent with our TAC hypothesis. TAC hypothesis confirmation needs a field demonstration that additional $CaCO_3$ precipitation is occurring each year, with more formed in summer than is decalcified in winter, in agreement with the $K_{sp}$ predictions. Obviously, the extent of this process will be a function of the seasonal variation in surface seawater temperature. We have expressed this according to the following Ψ (psi) equation, following a similar suggestion by Smith in 2014

(9) investigating whether the oceans were a net source or a sink for $CO_2$, showing variation with depth rather than seasonal variations in temperature.

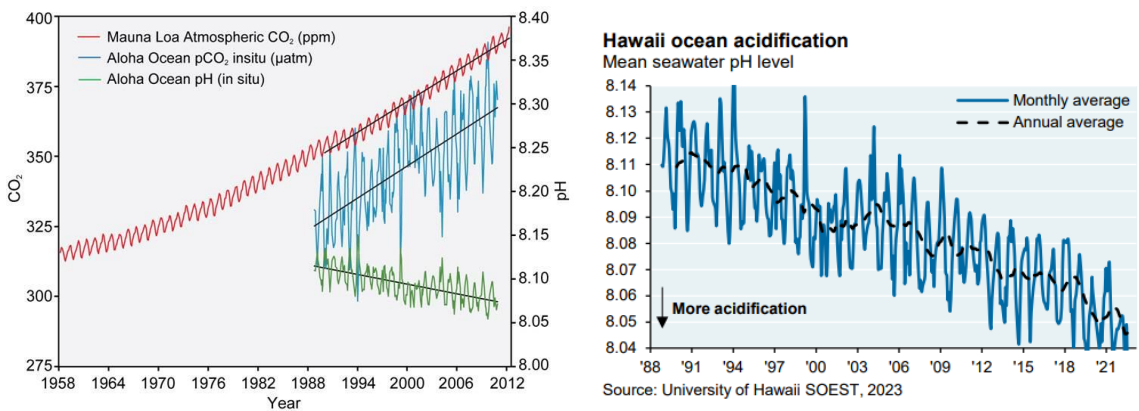

**Fig. 3.** NCCARF Plot of ALOHA Station Data, National Climate Change Research Facility, Australian Government and more recent data from the University of Hawaii.

The Ψ Equation

$$Ca^{2+} + 2(1 + \Psi)(HCO_3^- + H^+) <=> CaCO_3 + \Psi CO_3^{2-} + (1 + \Psi)(CH_2O + O_2) + 2(1 + \Psi)H^+$$

This fully reversible calcification equation moves towards acidification in summer and reverses to alkalinization in winter. The psi factor is a variable function of the range of seasonal changes in temperature. However, the greatest thermodynamic potential to emit $CO_2$ in seawater by acidification of bicarbonate ($HCO_3^-$) is when the pH value is lowest, the conversion of bicarbonate to $CO_2$ generating the greatest difference between $CO_2$ fugacity in seawater and that in air in mid-summer in northern hemisphere waters. The seasonal variation near Mauna Loa in atmospheric $pCO_2$ is about 6 ppmv whereas the long term increase in the Keeling curve year by year is one-third of this, about 2 ppmv suggesting that in spring and summer the $CO_2$ absorbed in about 2 ppmv, less than that emitted in autumn and winter. This increase is obvious by careful inspection of Fig. 1 and consistent with Fig. 4.

**WHOT Data**

The 18-year seasonal WHOT data of Chen et al. (*10*) shown in Fig. 4 confirms this seasonal variation in fugacity of $CO_2$ in the northern Pacific, even showing that value in surface seawater exceeds that in air for several months each year, speeding up transfer when the greatest differences occur between the phases at highest temperature. This is important too as a cause for increasing proportion of relative $^{12}CO_2$ in the atmosphere, decreasing the proportion of $^{13}CO_2$ even when it is increasing in the atmosphere. This is caused not by dilution of the atmosphere by fossil emissions low in $^{13}C$-content but by its predicted enrichment in air by concentration of $^{14}C$ isotopic carbon in seawater, a matter we will show elsewhere in a more detailed article. Obviously, there will be imbalances in photosynthesis and respiration with a local effect on $pCO_2$ levels, but to ascribe all the variation to terrestrial processes is unproven speculation. On the contrary, there is ample evidence that molar quantities of $CO_2$ per square meter of the sea surface are exchanged annually.

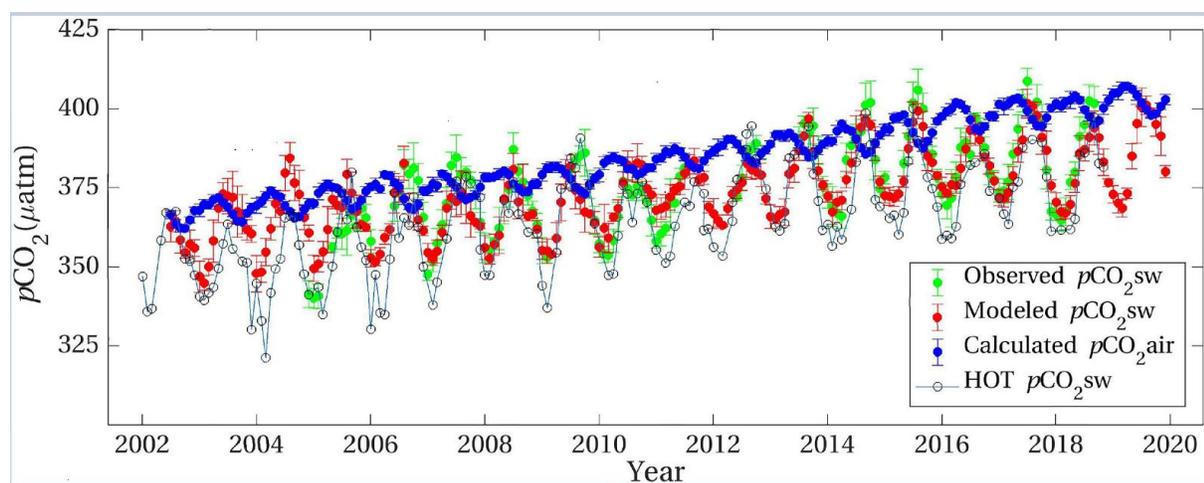

**Fig. 4.** Seasonal WHOT non-equilibrium oscillations in simultaneous $pCO_2$ measurements in seawater and air showing converse values supporting transfer of $CO_2$ to air in autumn through winter and, less strongly, from air to seawater in spring and summer. Peak and minimum seawater surface temperatures were 1-2 months after mid-summer and mid-winter with ranges of over 4 K, supporting the Figure 2 TAC model. Open access credit to Chen et al. (*10*)

The scale of the $pCO_2$ variation locally is correlated with the seasonal temperature variation. In general, variations are much smaller in the southern hemisphere, showing the greater depth of the mixing zone from which $CO_2$ exchange takes place, giving less surface variation in temperature seasonally. More research is needed.

**Discussion**

Limestone as a product of calcification is regarded as a biogeochemical product, given that phytoplankton and other marine organisms enhance its rate of production, if nutrients are available (*3*). In particular, the extracellular carbonic anhydrase apparently speeds the reversible dehydration of $CO_2$, forming bicarbonate and hydrated hydrogen ions ($H^+$) controlling pH. This article emphasizes that the reversible dehydration of $CO_2$ in surface seawater allowing transfer between aqueous and gaseous phases is most rate limiting of all, that carbonic anhydrase may even assist in autumn and winter, transferring $CO_2$ to the atmosphere.

However, all the calcification reactions shown in Fig. 2 also occur abiotically at a lower rate, increasing proportional to temperature. Discovery of $CaCO_3$ at the Phoenix landing site (*12*) was taken as evidence of liquid water on the surface of Mars, in geochemical terms at least. More recently (*13*), the identification of several carbonates such as siderite ($FeCO_3$) by the Curiosity rover also indicates that an active carbon cycle operated on Mars. Testing the TAC hypothesis may provide useful information for carbon cycling on other planets.

More significantly for managing climate change, if fossil fuel emissions are being largely absorbed by sequestration into the ocean surface and by 'greening' photosynthesis on land and in the ocean (*11*), the implications of this aspect of the TAC hypothesis for carbon-zero policies and renewable energy are profound. The thermal acid-calcification hypothesis predicts that global warming acidifies the ocean surface by increasing calcification causing $pCO_2$ to increase, independently of fossil emissions. Furthermore, this represents a striking illustration of the Le Chatelier principle, the carbon cycle on Earth responding intelligently to changing climate. The hope that carbon dioxide removal as sequestration (*14*), either biologically, chemically or geologically, by burial after capture, will prove futile.

In Figure 5 we present our current estimates of $CO_2$ cycling globally, per square meter of the Earth's surface. These numbers will be justified in more detail in an article in preparation. The testable TAC (thermal acid-calcification) hypothesis is a call for action in new scientific research. We recommend this be conducted by independent researchers rather than in IPCC-approved centers. As former Nature editor John Maddox critically claimed in his late 1990s book "*What Remains to Be Discovered*", research institutes free of political influence are needed for governments to consult with, on climate science. The uncertainty of the current IPCC paradigm regarding climate change and the role of fossil emissions of $CO_2$ in warming is large, lacking scientific evidence. A plausible alternative hypothesis offered here as the true cause of the increasing Keeling curve needs to be investigated urgently. This new model would still give predictively increasing emissions from the ocean in the complete absence of fossil fuel emissions because the acidification from calcification is purely a function of surface warming, from whatever cause.

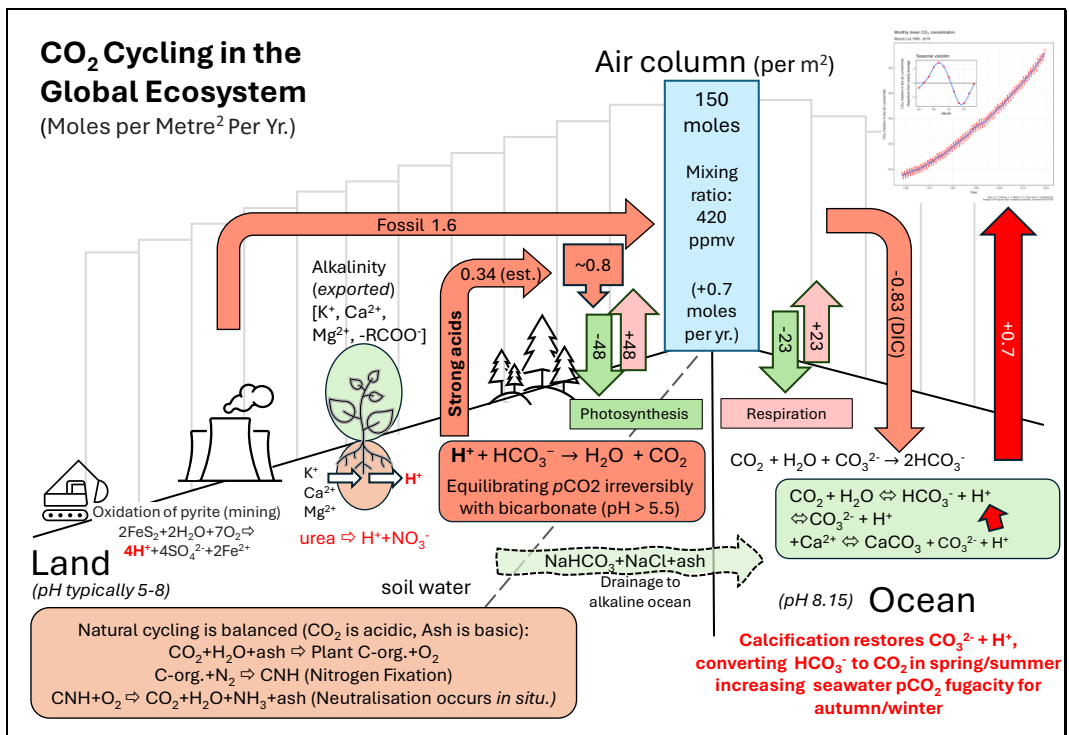

**Fig. 5.** Rates per square meter in global carbon cycling between land water, the atmosphere, and the ocean, illustrating the pH-acidification hypothesis. Emissions and absorptions shown are average moles per square meter of the Earth, for a mixing ratio of 420 ppmv in 2021 shown in the central column bridging land and ocean. The land acidification values are derived elsewhere (*11*), assuming photosynthesis is equal to respiration. The terrestrial area of Earth is $1.48 \times 10^{14}$ m$^{-2}$, the ocean's area is $3.62 \times 10^{14}$ m$^2$, $5.101 \times 10^{14}$ m$^2$ in total., represented as a mean value in the central air column.

In this article, we have deliberately left the causes of global warming for analysis elsewhere. In any case, this is likely to be a complex series of different cumulative peaks, both natural and human-caused caused. Even for the warming proposed by $CO_2$ in the IPCC radiative-convective model, a sensitivity factor of three must be applied, invoking twice as much warming from the extra water vapor, consequent on $CO_2$ raising temperature, a prediction not yet verified.